\documentclass[11pt]{article}
\usepackage{amssymb,amsmath,amsfonts}
\usepackage{graphicx}
\usepackage{graphics}
\usepackage{eepic,epsfig}

\makeatletter
\@addtoreset{equation}{section}

\makeatother

\begin{document}

\title{Scalar Casimir effect in a linearly expanding universe}
\author{A. A. Saharian$^{1}$ \thanks{%
E-mail: saharian@ysu.am }, \thinspace\ T. A. Petrosyan$^{1}$ \thanks{%
E-mail: tigran-petrosyan-1996@mail.ru }, \thinspace\ S. V. Abajyan$^{2}$
\thanks{%
E-mail: samvel.abajyan@mail.ru }, \thinspace\ B. B. Nersisyan$^{2}$ \thanks{%
E-mail: nerbab61@mail.ru } \vspace{0.3cm} \\
%EndAName
\textit{$^{1}$Department of Physics, Yerevan State University,}\\
\textit{1 Alex Manoogian Street, 0025 Yerevan, Armenia}\vspace{0.3cm}\\
\textit{$^2$Armenian State Pedagogical University,}\\
\textit{13 Khandjyan Street, 0010 Yerevan, Armenia}}
\maketitle

\begin{abstract}
We investigate quantum vacuum effects for a massive scalar field, induced by
two planar boundaries in background of a linearly expanding spatially flat
Friedmann-Robertson-Walker spacetime for an arbitrary number of spatial
dimensions. For the Robin boundary conditions and for general curvature
coupling parameter, a complete set of mode functions is presented and the
related Hadamard function is evaluated. The results are specified for the
most important special cases of the adiabatic and conformal vacuum states.
The vacuum expectation values of the field squared and of the
energy-momentum tensor are investigated for a massive conformally coupled
field. The vacuum energy-momentum tensor, in addition to the diagonal
components, has nonzero off-diagonal component describing energy flux along
the direction perpendicular to the plates. The influence of the
gravitational field on the local characteristics of the vacuum state is
essential at distances from the boundaries larger than the curvature radius
of the background spacetime. In contrast to the Minkowskian bulk, at large
distances the boundary-induced expectation values follow as power law for
both massless and massive fields. Another difference is that the Casimir
forces acting on the separate plates do not coincide if the corresponding
Robin coefficients are different. At large separations between the plates
the decay of the forces is power law. We show that during the cosmological
expansion the forces may change the sign.
\end{abstract}

\bigskip

PACS numbers: 04.62.+v, 03.70.+k, 98.80.-k

\bigskip

\section{Introduction}

The Casimir effect (for reviews see \cite{Most97}) is among the most
interesting quantum field-theoretical effects having a macroscopic
manifestation. The effect arises as a consequence of the modification of the
spectrum for the vacuum fluctuations caused by the imposition of boundary
conditions on the operator of a quantum field. As a result of that, the
vacuum expectation values (VEVs) of physical observables are shifted by an
amount depending on the bulk and boundary geometries and on the specific
boundary conditions. In particular, vacuum forces arise acting on the
constraining boundaries. For the quantum electromagnetic field these forces
have been measured in a large number of experiments.

An interesting topic in the investigations of the Casimir effect is the
dependence of the vacuum characteristics on the geometry of the background
spacetime. Exact results are obtained for highly symmetric geometries only.
In particular, the consideration of quantum effects in cosmological
backgrounds has attracted a great deal of attention (see, for instance, \cite%
{Birr82B}). The boundary conditions on fields in cosmological models may
have different physical origins. They can be caused by nontrivial spatial
topology (for example, in Kaluza-Klein type models with extra dimensions),
by the presence of coexisting phases \cite{Bell14b}, by topological defects,
or by branes in the scenarios of the braneworld type. All these sources of
boundary conditions give arise additional contributions to the physical
characteristics of the vacuum state. In our previous research on the Casimir
effect on curved backgrounds we have considered various bulk and boundary
geometries. Among the most popular geometries is the de Sitter (dS)
spacetime. In particular, the VEVs for planar boundaries on this background
have been discussed in \cite{Seta01,Saha09,Eliz10} and \cite{Saha14} for
scalar and electromagnetic fields, respectively. The corresponding Casimir
densities for spherical and cylindrical boundaries were investigated as well
\cite{Milt12,Saha15} (for the Casimir effect on background of the anti-de
Sitter (AdS) spacetime see references given in \cite{Eliz13}). The VEVs of
the electric and magnetic field squared and of the energy-momentum tensor
for the electromagnetic field, induced by a single and two parallel
conducting plates in spatially flat Friedmann-Robertson-Walker (FRW)
universes with a power-law scale factor have been evaluated in \cite{Bell13}%
. The quantum vacuum effects for a scalar field in the presence of by planar
boundaries for a spatially flat bulk with a general scale factor are studied
in \cite{Beze17}.

In the present paper we consider the scalar Casimir densities and forces for
the geometry of two parallel plates in background of a linearly expanding
spatially flat $(D+1)$-dimensional cosmological model. The latter is among
the simplest cosmological backgrounds allowed by string theories \cite%
{Anto89}. The corresponding dilaton field behaves as $\Phi =(1-D)\ln t+%
\mathrm{const}$. Various aspects of quantum field theory in a linearly
expanding universe have been discussed in \cite{Dise74}-\cite{Toll02}. Among
the most interesting effects allowing a comprehensive study are the vacuum
polarization and particle production by the time-dependent gravitational
field. Though in our consideration the presence of the boundaries breaks the
homogeneity of the background geometry, we will show that the corresponding
Casimir problem is still exactly solvable for a class of Robin boundary
conditions with the coefficients (in general, different on separate plates)
proportional to the scale factor. These coefficients can be interpreted in
terms of the finite penetration length of the field to the boundary. For a
scalar field with general curvature coupling parameter, the corresponding
Casimir problems on the dS, Minkowski and AdS bulks have been considered in
\cite{Eliz10,RS02,Saha05}, respectively.

The organization of the paper is as follows. In the next section, the bulk
and boundary geometries and the field content are specified. For the
evaluation of the VEVs we use the summation over a complete set of scalar
modes and the corresponding mode functions are presented in section \ref%
{sec:ScModes}. In section \ref{sec:Asymp} we discuss the asymptotics of the
mode functions and the most important special cases corresponding to the
adiabatic and conformal vacuum states. In section \ref{sec:TwoPoint}, a
general expression for the Hadamard function is obtained and then it is
further transformed for the case of a conformally coupled scalar filed
prepared in the conformal vacuum. The VEVs of the field squared and of the
energy-momentum tensor for this special case are investigated in sections %
\ref{sec:phi2} and \ref{sec:EMT}. The Casimir forces are studied in section %
\ref{sec:Force}. And, finally, the main results of the paper are summarized
in section \ref{sec:Conc}.

\section{Problem setup}

\label{sec:Problem}

As a background geometry we consider a linearly expanding $(D+1)$%
-dimensional universe described by the line element%
\begin{equation}
ds^{2}=dt^{2}-a^{2}(t)d\mathbf{x}^{2},\;a(t)=bt,  \label{ds2}
\end{equation}%
with spatial coordinates $\mathbf{x}=(x^{1},x^{2},\ldots ,x^{D})$. In (\ref%
{ds2}), $0\leqslant t<\infty $ and $b>0$ is a constant having dimension of
inverse length. Introducing a conformal time $\eta $, $-\infty <\eta
<+\infty $, in accordance with%
\begin{equation}
t=e^{b\eta }/b,  \label{teta}
\end{equation}%
the line element is written in explicitly conformally flat form%
\begin{equation}
ds^{2}=a^{2}(\eta )\left( d\eta ^{2}-d\mathbf{x}^{2}\right) ,\;a(\eta
)=e^{b\eta }.  \label{ds2c}
\end{equation}%
In what follows we will work in the spacetime coordinate system $(\eta ,%
\mathbf{x})$. In these coordinates, the Ricci scalar, $R$, and the nonzero
components of the Ricci tensor, $R_{\mu \nu }$, are given by the expressions%
\begin{equation}
R=D(D-1)b^{2}e^{-2b\eta },\;R_{00}=0,\;R_{ik}=-(D-1)b^{2}\delta _{ik},
\label{RRik}
\end{equation}%
with $i,k=1,2,\ldots ,D$. From the Einstein equations for the corresponding
energy density $\varepsilon $ and the pressure $p$ one has%
\begin{equation}
\varepsilon =\frac{D(D-1)}{16\pi Gt^{2}},\;p=-\frac{D-2}{D}\varepsilon ,
\label{epsp}
\end{equation}%
where $G$ is the Newton gravitational constant. For $D=1$, the geometry we
have described is flat and coincides with the $(1+1)$-dimensional Milne
universe.

Having specified the background geometry let us turn to the field content of
the problem. We will consider a massive scalar field $\varphi (x)$ with the
curvature coupling parameter $\xi $. The corresponding field equation reads
\begin{equation}
(\nabla _{\mu }\nabla ^{\mu }+m^{2}+\xi R)\varphi =0,  \label{Feq}
\end{equation}%
where $\nabla _{\mu }$ stands for the covariant derivative operator. Here we
are interested in the effects on the scalar vacuum induced by codimension
one flat boundaries (plates) located at $x^{D}\equiv z=z_{1}$ and $z=z_{2}$,
$z_{2}>z_{1}$. On the plate $z=z_{j}$, $j=1,2$, the field operator is
constrained by the boundary condition $(1+\beta _{j}^{\prime }n_{j}^{\mu
}\nabla _{\mu })\varphi =0$, with $n_{j}^{\mu }$ being the normal to the
boundary obeying the relation $n_{j\mu }n_{j}^{\mu }=-1$. The boundary
conditions considered are of the Robin type and generalize the Dirichlet ($%
\beta _{j}^{\prime }=0$) and Neumann ($\beta _{j}^{\prime }=\infty $)
boundary conditions. In the regions $z<z_{1}$ and $z>z_{2}$ for the normal
one has $n_{1}^{\mu }=-\delta _{D}^{\mu }e^{-b\eta }$ and $n_{2}^{\mu
}=\delta _{D}^{\mu }e^{-b\eta }$, respectively. For the region $%
z_{1}\leqslant z\leqslant z_{2}$ the normal is given by $n_{j}^{\mu
}=(-1)^{j-1}\delta _{D}^{\mu }e^{-b\eta }$ for $j=1,2$. The coefficients $%
\beta _{j}^{\prime }$ have the dimension of length and in some problems
characterize the penetration depth of the field. In what follows a special
case will be considered with the Robin coefficients $\beta _{j}^{\prime
}=\beta _{j}e^{b\eta }=\beta _{j}bt$, where $\beta _{j}$, $j=1,2$, are
constants (the penetration length scales proportional to the scale factor).
In this case, the boundary conditions in the region between the plates take
the form
\begin{equation}
\lbrack 1+(-1)^{j-1}\beta _{j}\partial _{z}]\phi =0,  \label{7}
\end{equation}%
for $\;z=z_{j}$.

The boundary conditions imposed on the field modify the spectrum of
zero-point fluctuations and, as a consequence, the VEVs of physical
observables are shifted. The VEVs are expressed in terms of the two-point
functions. The latter can be presented in the form of the sum over a
complete set of solutions to the field equation (\ref{Feq}) obeying the
boundary conditions. These solutions for the problem under consideration are
specified in the next section.

\section{Complete set of modes}

\label{sec:ScModes}

The background geometry is flat and for the complete set of scalar modes in
the region between the plates, $z_{1}\leqslant z\leqslant z_{2}$, the
dependence on the spatial coordinates can be taken similar to that for
plates in the Minkowski bulk:
\begin{equation}
\varphi (x)=Cf(\eta )e^{i\mathbf{k}\cdot \mathbf{x}_{\parallel }}\cos \left[
\lambda \left( z-z_{j}\right) +\alpha _{j}(\lambda )\right] \ ,
\label{ansatz}
\end{equation}%
where $\mathbf{k}=(k^{1},k^{2},\ldots ,k^{D-1})$, $\mathbf{x}_{\parallel
}=(x^{1},x^{2},\ldots ,x^{D-1})$, the function $\alpha _{j}(\lambda )$ is
defined as
\begin{equation}
e^{2i\alpha _{j}(\lambda )}=\frac{i\lambda \beta _{j}+(-1)^{j}}{i\lambda
\beta _{j}-(-1)^{j}},  \label{ealfj}
\end{equation}%
and $C$ is the normalization constant. The modes (\ref{ansatz}) obey the
boundary condition on the plate $z=z_{j}$. From the boundary condition on
the second plate it follows that the eigenvalues of the quantum number $%
\lambda $ are roots of the equation
\begin{equation}
(1-b_{1}b_{2}u^{2})\sin u-(b_{1}+b_{2})u\cos u\ =0,  \label{rest}
\end{equation}%
with the notations%
\begin{equation}
u=\lambda z_{0},\;b_{j}=\beta _{j}/z_{0},\;z_{0}=z_{2}-z_{1}.  \label{bj}
\end{equation}%
The equation (\ref{rest}) coincides with the eigenvalue
equation for plates in the Minkowski bulk \cite{RS02}. We will denote the
roots of the transcendental equation (\ref{rest}) by $u=u_{n}$, $%
n=1,2,\ldots $. For the eigenvalues of the quantum number $\lambda $ one has
$\lambda =\lambda _{n}=u_{n}/z_{0}$. In the discussion below we will assume
the values of the parameters $b_{j}$ for which all the roots $u_{n}$ are
real (for possible purely imaginary roots see \cite{RS02}). In particular,
this is the case for $\beta _{j}\leqslant 0$.

In order to determine the function $f(\eta )$, we substitute (\ref{ansatz})
into the field equation (\ref{Feq}). This leads to the equation
\begin{equation}
f^{\prime \prime }(\eta )+(D-1)bf^{\prime }(\eta )+\left[ \gamma ^{2}+\xi
D(D-1)b^{2}+m^{2}e^{2b\eta }\right] f(\eta )=0,  \label{f1}
\end{equation}%
where $\gamma =\sqrt{\lambda ^{2}+{k}^{2}}$, $k=|\mathbf{k}|$, and the prime
stands for the derivative with respect to $\eta $. The solution of this
equation is expressed in terms of cylindrical functions as%
\begin{equation}
f(\eta )=(bt)^{(1-D)/2}\left[ w_{1}e^{-\nu \pi /2}H_{i\nu
}^{(1)}(mt)+w_{2}e^{\nu \pi /2}H_{i\nu }^{(2)}(mt)\right] ,  \label{fM}
\end{equation}%
with $H_{i\nu }^{(l)}(x)$, $l=1,2$, being the Hankel functions,
\begin{equation}
\nu =\sqrt{\gamma ^{2}b^{-2}+\xi D(D-1)-\left( D-1\right) ^{2}/4},
\label{nu}
\end{equation}%
and $t$ is expressed in terms of the conformal time $\eta $ as (\ref{teta}).
The function $\nu =\nu (\gamma )$ can be either positive or purely
imaginary. In (\ref{fM}), the coefficients $w_{1}$ and $w_{2}$, in general,
can be functions of $\gamma $. The factors $e^{\pm \nu \pi /2}$ are
extracted for the further convenience. In what follows we will assume that
the function $f(\eta )$ is normalized by the condition%
\begin{equation}
f(\eta )f^{\ast \prime }(\eta )-f^{\ast }(\eta )f^{\prime }(\eta
)=ie^{\left( 1-D\right) b\eta },  \label{cond2}
\end{equation}%
where the star stands for the complex conjugate. Substituting (\ref{fM}) and
using the Wronskian relation for the Hankel functions, one gets the relation
between the coefficients%
\begin{equation}
\left\vert w_{2}\right\vert ^{2}-\left\vert w_{1}\right\vert ^{2}=\frac{\pi
}{4b}.  \label{relc12}
\end{equation}

We can write the solution (\ref{fM}) in terms of the Bessel function $%
J_{i\nu }(z)$:
\begin{equation}
f(\eta )=(bt)^{(1-D)/2}\left[ d_{1}J_{-i\nu }(mt)+d_{2}J_{i\nu }(mt)\right] ,
\label{fJ}
\end{equation}%
where, again, $t$ is given by (\ref{teta}). The coefficients $d_{1}$ and $%
d_{2}$ are related to the previous ones by the formulas%
\begin{equation}
d_{1}=\frac{w_{2}e^{\nu \pi /2}-w_{1}e^{-\nu \pi /2}}{\sinh \left( \nu \pi
\right) },\quad d_{2}=\frac{w_{1}e^{\nu \pi /2}-w_{2}e^{-\nu \pi /2}}{\sinh
\left( \nu \pi \right) },  \label{d12}
\end{equation}%
and the vice versa%
\begin{equation}
w_{1}=\frac{e^{-\nu \pi /2}d_{1}+e^{\nu \pi /2}d_{2}}{2},\quad w_{2}=\frac{%
e^{\nu \pi /2}d_{1}+e^{-\nu \pi /2}d_{2}}{2}.  \label{c12}
\end{equation}%
From (\ref{relc12}) we obtain the following relation between the new
coefficients
\begin{equation}
\left( |d_{1}|^{2}-|d_{2}|^{2}\right) \sinh \left[ (\nu +\nu ^{\ast })\frac{%
\pi }{2}\right] +\left( d_{1}d_{2}^{\ast }-d_{1}^{\ast }d_{2}\right) \sinh %
\left[ (\nu -\nu ^{\ast })\frac{\pi }{2}\right] =\frac{\pi }{2b}.
\label{reld12}
\end{equation}

So, for the complete set of solutions one has $\{\varphi _{n\mathbf{k}%
}^{(+)}(x),\varphi _{n\mathbf{k}}^{(-)}(x)=\varphi _{n\mathbf{k}}^{(+)\ast
}(x)\}$, with
\begin{equation}
\varphi _{n\mathbf{k}}^{(+)}(x)=Cf(\eta ,\gamma _{n})e^{i\mathbf{k}\cdot
\mathbf{x}_{\parallel }}\cos \left[ \lambda _{n}\left( z-z_{j}\right)
+\alpha _{j}(\lambda _{n})\right] \ ,  \label{phi+}
\end{equation}%
where we have explicitly displayed the dependence of the function $f$ on $%
\gamma _{n}=\sqrt{\lambda _{n}^{2}+{k}^{2}}$. From the orthonormalization
condition of the scalar modes, for the coefficient $C$ one gets%
\begin{equation}
\left\vert C\right\vert ^{2}=\frac{2}{\left( 2\pi \right) ^{D-1}z_{0}c_{n}}%
,\;c_{n}=1+\frac{\sin u_{n}}{u_{n}}\cos [u_{n}+2\tilde{\alpha}_{j}(u_{n})]\ ,
\label{Csig}
\end{equation}%
where the function $\tilde{\alpha}_{j}(u)$ is defined as $e^{2i\tilde{\alpha}%
_{j}(u)}=(ub_{j}+i)/(ub_{j}-i)$, with $j=1,2$. Note that the mode functions (%
\ref{phi+}) are not completely fixed by the normalization condition: one of
the coefficients in the representations (\ref{fM}) or (\ref{fJ}) remains
arbitrary. It is determined by the choice of the vacuum state $|0\rangle $.
For example, an additional condition could be the requirement of the smooth
transition to the standard Minkowskian vacuum in the limit of slow expansion
(see below). The scalar mode functions for a conformally coupled scalar
field in the boundary-free geometry for the special case $D=3$ have been
discussed in \cite{Chit77,Nari78,Char81,Azum82,Calz83}. Another special case
$D=1$ is considered in \cite{Birr82B}.

\section{Asymptotics of the mode functions and the vacuum states}

\label{sec:Asymp}

Here we consider the most important special cases of the mode functions
realizing the adiabatic and conformal vacua.

\subsection{Adiabatic vacuum}

First let us consider the Minkowskian limit. As seen from (\ref{ds2c}), in
this limit $b\rightarrow 0$ for fixed $\eta $ and, consequently, $mt\approx
m/b+m\eta \gg 1$. For the function $\nu $ one has $\nu \approx \gamma /b$.
This means that both the argument and the absolute value of the order for
the Hankel functions in (\ref{fM}) are large. By using the uniform
asymptotic expansions for the Hankel functions one gets
\begin{equation}
f(\eta ,\gamma )\approx \sqrt{\frac{2b}{\pi \omega }}\left[ w_{1}e^{i\nu \xi
(m/\gamma )-i\pi /4}e^{i\omega \eta }+w_{2}e^{-i\nu \xi (m/\gamma )+i\pi
/4}e^{-i\omega \eta }\right] ,  \label{fMM}
\end{equation}%
where $\omega =\sqrt{\gamma ^{2}+m^{2}}$ and
\begin{equation}
\xi (u)=\sqrt{1+u^{2}}+\ln \left( \frac{u}{1+\sqrt{1+u^{2}}}\right) .
\label{ksi}
\end{equation}%
For the coefficients we have the relation (\ref{relc12}).

From (\ref{fMM}) it follows that the state under consideration is reduced to
the Minkowskian vacuum if $w_{1}=0$. The vacuum state obeying this property
is called an adiabatic vacuum. For this vacuum $\left\vert w_{2}\right\vert
^{2}=\pi /(4b)$ and, in the Minkowskian limit, the modes $\varphi _{\sigma
}^{(+)}(x)$ coincide (up to a phase) with the positive-energy modes in the
Minkowski spacetime. Hence, for the modes realizing the adiabatic vacuum one
has%
\begin{equation}
f(\eta ,\gamma )=f_{(A)}(\eta ,\gamma )=w_{2}e^{\nu \pi
/2}(bt)^{(1-D)/2}H_{i\nu }^{(2)}(mt),\;\left\vert w_{2}\right\vert ^{2}=%
\frac{\pi }{4b}.  \label{fMM2}
\end{equation}%
The corresponding state will be denoted as $|0_{A}\rangle $.

\subsection{Conformal vacuum}

Consider a conformally coupled massless field for which $\xi =(D-1)/(4D)$
and, hence, $\nu =\gamma /b$. By using the asymptotic expression for the
Bessel function for small arguments \cite{Abra72}, from (\ref{fJ}) in the
limit $m\rightarrow 0$ one gets%
\begin{equation}
f(\eta ,\gamma )=e^{(1-D)b\eta /2}\left[ \frac{d_{1}e^{-i\nu \ln (m/2b)}}{%
\Gamma \left( 1-i\nu \right) }e^{-i\gamma \eta }+\frac{d_{2}e^{i\nu \ln
(m/2b)}}{\Gamma \left( 1+i\nu \right) }e^{i\gamma \eta }\right] ,
\label{fJC}
\end{equation}%
where $\Gamma (x)$ is the gamma function. The corresponding modes are
conformally related to the positive-energy mode functions in the Minkowski
bulk if $d_{2}=0$. This correspond to the following relation for the
coefficients $w_{1,2}$:
\begin{equation}
w_{2}=w_{1}e^{\nu \pi }.  \label{w2c}
\end{equation}%
From (\ref{reld12}) one finds
\begin{equation}
|d_{1}|^{2}=\frac{\pi }{2b\sinh \left( \gamma \pi /b\right) }.
\label{reld12C}
\end{equation}%
By taking into account that%
\begin{equation}
\Gamma \left( 1+i\gamma /b\right) \Gamma \left( 1-i\gamma /b\right) =\frac{%
\gamma /b}{\sinh (\pi \gamma /b)},  \label{GamRel}
\end{equation}%
for the corresponding modes, up to a phase, one gets%
\begin{equation}
f(\eta ,\gamma )=e^{(1-D)b\eta /2}\frac{e^{-i\gamma \eta }}{\sqrt{2\gamma }}.
\label{fclim}
\end{equation}%
Note that the corresponding function in the positive-energy modes for the
Minkowski bulk is given by $e^{-i\gamma \eta }/\sqrt{2\gamma }$.

The vacuum state defined by the mode functions with $d_{2}=0$ is called a
conformal vacuum and we will denote it as $|0_{C}\rangle $. For the
corresponding modes we get%
\begin{equation}
f(\eta ,\gamma )=f_{(C)}(\eta ,\gamma )=d_{1}(bt)^{(1-D)/2}J_{-i\nu }(mt),
\label{fJC2}
\end{equation}%
where%
\begin{equation}
|d_{1}|^{2}\sinh \left[ (\nu +\nu ^{\ast })\frac{\pi }{2}\right] =\frac{\pi
}{2b}.  \label{reld12C1}
\end{equation}%
From here it follows that the conformal vacuum is physically realizable for
real values of $\nu $ only and the mode functions are given by (\ref{fJC2})
with%
\begin{equation}
|d_{1}|^{2}=\frac{\pi }{2b\sinh (\nu \pi )}.  \label{d1C}
\end{equation}%
Note that the mode functions for the adiabatic and conformal vacua in the
Milne universe with $D=1$ and in the absence of plates have been discussed
in \cite{Birr82B} (see also references given therein).

For the modes realizing the conformal and adiabatic vacua one has the
relation%
\begin{equation}
\varphi _{(C)n^{\prime }\mathbf{k}^{\prime }}^{(+)}(x)=\sum_{n}\int d\mathbf{%
k}\,\left[ \alpha _{n^{\prime }\mathbf{k}^{\prime },n\mathbf{k}}\varphi
_{(A)n\mathbf{k}}^{(+)}(x)+\beta _{n^{\prime }\mathbf{k}^{\prime },n\mathbf{k%
}}\varphi _{(A)n\mathbf{k}}^{(-)}(x)\right] ,  \label{Bog1}
\end{equation}%
where the mode functions $\varphi _{(C)n\mathbf{k}}^{(+)}(x)$ and $\varphi
_{(A)n\mathbf{k}}^{(+)}(x)$ are given by (\ref{phi+}) with $f(\eta ,\gamma
)=f_{(C)}(\eta ,\gamma )$ and $f(\eta ,\gamma )=f_{(A)}(\eta ,\gamma )$,
respectively. For the Bogoliubov coefficients in (\ref{Bog1}) we get%
\begin{equation}
\alpha _{n^{\prime }\mathbf{k}^{\prime },n\mathbf{k}}=\frac{e^{\nu \pi
/2}\delta _{n^{\prime }n}\delta (\mathbf{k}^{\prime }-\mathbf{k})}{\sqrt{%
2\sinh (\nu \pi )}},\;\beta _{n^{\prime }\mathbf{k}^{\prime },n\mathbf{k}}=%
\frac{e^{-\nu \pi /2}\delta _{n^{\prime }n}\delta (\mathbf{k}^{\prime }+%
\mathbf{k})}{\sqrt{2\sinh (\nu \pi )}}.  \label{Bog2}
\end{equation}%
One has $\beta _{n^{\prime }\mathbf{k}^{\prime },n\mathbf{k}}\neq 0$ and the
conformal vacuum contains particles defined by using the adiabatic modes.
For the mean number of particles (per unit volume along the directions
parallel to the plates) with quantum numbers $(n,\mathbf{k})$ we find $%
\langle 0_{C}|N_{(A)n\mathbf{k}}|0_{C}\rangle =1/(e^{\nu \pi }-1)$.

\section{Two-point function}

\label{sec:TwoPoint}

As a two-point function we will consider the Hadamard function defined as
the VEV $G(x,x^{\prime })=$ $\langle 0|\phi (x)\phi (x^{\prime })+\phi
(x^{\prime })\phi (x)|0\rangle $. Expanding the field operator in terms of
the complete set of mode functions and using the commutation relations for
the annihilation and creation operators, the following mode-sum formula is
obtained:%
\begin{equation}
G(x,x^{\prime })=\int d\mathbf{k}\,\sum_{n=1}^{\infty }\sum_{s=\pm }\varphi
_{n\mathbf{k}}^{(s)}(x)\varphi _{n\mathbf{k}}^{(s)\ast }(x^{\prime }),
\label{Had}
\end{equation}%
with the mode functions given by (\ref{phi+}).

\subsection{General expression}

Substituting the mode functions (\ref{phi+}) one gets the representation%
\begin{eqnarray}
G(x,x^{\prime }) &=&\frac{\left( tt^{\prime }\right) ^{(1-D)/2}}{\left( 2\pi
b\right) ^{D-1}z_{0}}\int d\mathbf{k}\,e^{i\mathbf{k}\cdot \Delta \mathbf{x}%
_{\parallel }}\sum_{n=1}^{\infty }\frac{W(t,t^{\prime },\gamma _{n})}{c_{n}}
\notag \\
&&\times \left\{ \cos \left( \lambda _{n}\Delta z\right) +\cos [\lambda
_{n}\left( z+z^{\prime }-2z_{j}\right) +2\alpha _{j}(\lambda _{n})]\right\} ,
\label{Had1}
\end{eqnarray}%
where $\Delta \mathbf{x}_{\parallel }=\mathbf{x}_{\parallel }-\mathbf{x}%
_{\parallel }^{\prime }$, $\Delta z=z-z^{\prime }$ and%
\begin{eqnarray}
W(t,t^{\prime },\gamma ) &=&\left[ |w_{1}|^{2}+|w_{2}|^{2}\right] \left[
H_{i\nu }^{(1)}(mt)H_{i\nu }^{(2)}(mt^{\prime })+H_{i\nu }^{(1)}(mt^{\prime
})H_{i\nu }^{(2)}(mt)\right]  \notag \\
&&+2w_{1}w_{2}^{\ast }e^{-\nu \pi }H_{i\nu }^{(1)}(mt)H_{i\nu
}^{(1)}(mt^{\prime })+2w_{1}^{\ast }w_{2}e^{\nu \pi }H_{i\nu
}^{(2)}(mt)H_{i\nu }^{(2)}(mt^{\prime }).  \label{wg}
\end{eqnarray}%
For the adiabatic vacuum $w_{1}=0$ and we find%
\begin{equation}
W(t,t^{\prime },\gamma )=\frac{\pi }{4b}\left[ H_{i\nu }^{(1)}(mt)H_{i\nu
}^{(2)}(mt^{\prime })+H_{i\nu }^{(1)}(mt^{\prime })H_{i\nu }^{(2)}(mt)\right]
.  \label{wa}
\end{equation}%
For the conformal vacuum
\begin{equation}
w_{1}=e^{-\nu \pi /2}d_{1}/2,\quad w_{2}=e^{\nu \pi /2}d_{1}/2,  \label{w12c}
\end{equation}%
and the function $W(t,t^{\prime },\gamma )$ is given by
\begin{equation}
W(t,t^{\prime },\gamma )=\pi \frac{J_{-i\nu }(mt)J_{i\nu }(mt^{\prime
})+J_{i\nu }(mt)J_{-i\nu }(mt^{\prime })}{2b\sinh (\nu \pi )}.  \label{wc}
\end{equation}%
Recall that for the conformal vacuum $\nu $ should be real. Note that in the
case of the adiabatic vacuum the function $W(t,t^{\prime },\gamma )$ is an
even function of $\nu $, whereas for the conformal vacuum it is an odd
function of $\nu $.

\subsection{Hadamard function for the conformal vacuum}

In the further discussion we will consider the conformal vacuum and a
conformally coupled scalar field. For the latter $\xi =(D-1)/4D$ and $\nu
=\gamma _{n}/b$. Hence, the function in the expression (\ref{Had1}) of the
Hadamard function takes the form%
\begin{equation}
W(t,t^{\prime },\gamma )=\pi \frac{J_{-i\gamma /b}(mt)J_{i\gamma
/b}(mt^{\prime })+J_{i\gamma /b}(mt)J_{-i\gamma /b}(mt^{\prime })}{2b\sinh
(\pi \gamma /b)}.  \label{wcc}
\end{equation}%
The Hadamard function (\ref{Had1}) with (\ref{wcc}) is further transformed
by using a variant of the generalized Abel-Plana summation formula \cite%
{RS02,SahaBook}:%
\begin{equation}
\sum_{n=1}^{\infty }\frac{g(u_{n})}{c_{n}}=-\frac{g(0)/2}{1-b_{2}-b_{1}}+%
\frac{1}{\pi }\int_{0}^{\infty }du\,g(u)+\frac{i}{\pi }\int_{0}^{\infty }du%
\frac{g(iu)-g(-iu)}{c_{1}(u)c_{2}(u)e^{2u}-1},  \label{sumfor}
\end{equation}%
with the notation%
\begin{equation}
c_{j}(u)=\frac{b_{j}u-1}{b_{j}u+1}.  \label{cj}
\end{equation}%
In this formula we take the function
\begin{equation}
g(u)=\left\{ \cos \left( u\Delta z/z_{0}\right) +\cos [u\left( z+z^{\prime
}-2z_{j}\right) /z_{0}+2\alpha _{j}(u/z_{0})]\right\} W(t,t^{\prime },\sqrt{%
u^{2}/z_{0}^{2}+k^{2}}).  \label{gu}
\end{equation}%
For the latter one has $g(iu)-g(-iu)=0$ for $u<kz_{0}$ and $%
g(iu)-g(-iu)=2g(iu)$ for $u>kz_{0}$.

In deriving the summation formula (\ref{sumfor}) from the generalized
Abel-Plana formula in \cite{RS02,SahaBook}, it was assumed that the function
$g(u)$ is analytic in the right half of the complex plane Re$\,u\geqslant 0$
and obeys the condition $|g(u)|<\epsilon (x)e^{c|y|}$ for $|u|\rightarrow
\infty $, where $u=x+iy$, $c<2$, and $\epsilon (x)\rightarrow 0$ for $%
x\rightarrow \infty $. The function (\ref{gu}) obeys these conditions except
the analyticity on the imaginary axis Re$\,u=0$: the function $g(u)$ has
simple poles $u=\pm iy_{l}$, $l=1,2,\ldots $, with
\begin{equation}
y_{l}=z_{0}\sqrt{k^{2}+l^{2}b^{2}},  \label{yl}
\end{equation}%
coming from the zeros of the denominator in (\ref{wcc}). Note that in the
discussion of boundary-induced vacuum quantum effects on the FRW background,
presented in \cite{Beze17}, it was assumed that the corresponding integrand
is analytic in the right half-plane. Hence, the expressions for the vacuum
characteristics in the problem under consideration cannot be directly
obtained from the results in \cite{Beze17}.

In the derivation of the summation formula (\ref{sumfor}) from the
generalized Abel-Plana formula (see \cite{SahaBook}) the poles $\pm iy_{l}$
should be excluded by small semicircles $C_{\rho }^{\pm }$ with radius $\rho
$ on the right half-plane, with the subsequent limiting transition $\rho
\rightarrow 0$. The contributions of the integrals along these semicircles
to the right-hand side of (\ref{sumfor}) is expressed as
\begin{equation}
\frac{1}{2}\sum_{j=+,-}\int_{C_{\rho }^{j}}du\frac{h_{j}(u)}{\sinh (\pi
\sqrt{u^{2}/z_{0}^{2}+{k}^{2}}/b)},  \label{PolContr}
\end{equation}%
where%
\begin{eqnarray}
h_{\pm }(u) &=&\frac{\pi i}{2b}(b_{1}u\pm i)(b_{2}u\pm i)e^{\pm iu}  \notag
\\
&&\times \frac{J_{-i\gamma /b}(mt)J_{i\gamma /b}(mt^{\prime })+J_{i\gamma
/b}(mt)J_{-i\gamma /b}(mt^{\prime })}{(1-b_{1}b_{2}u^{2})\sin
u-(b_{1}+b_{2})u\cos u}.  \label{hpm}
\end{eqnarray}%
For the separate integrals one has
\begin{equation}
\int_{C_{\rho }^{\pm }}du\frac{h_{\pm }(u)}{\sinh (\pi \gamma /b)}%
=ilb^{2}z_{0}^{2}\frac{h_{\pm }(\pm iy_{l})}{(-1)^{l}y_{l}}.  \label{IntC}
\end{equation}%
Now, it can be seen that $h_{-}(-iw_{l})=-h_{+}(iw_{l})$ and, hence, in (\ref%
{PolContr}) the contributions coming from the poles $iy_{l}$ and $-iy_{l}$
cancel each other. From here we conclude that the summation formula (\ref%
{sumfor}) is valid for the function (\ref{gu}) if the last integral in the
right-hand side is understood in the sense of the principal value.

Applying the summation formula (\ref{sumfor}) to the series in (\ref{Had1})
and introducing the function%
\begin{equation}
V(t,t^{\prime },\chi )=\frac{J_{\chi }(mt)J_{-\chi }(mt^{\prime })+J_{-\chi
}(mt)J_{\chi }(mt^{\prime })}{\sin (\pi \chi )},  \label{Vxi}
\end{equation}%
the Hadamard function is presented in the form%
\begin{eqnarray}
G(x,x^{\prime }) &=&G_{j}(x,x^{\prime })+\frac{\left( b^{2}tt^{\prime
}\right) ^{(1-D)/2}}{\left( 2\pi \right) ^{D-1}bz_{0}}\int d\mathbf{k}%
\,\int_{kz_{0}}^{\infty }du\frac{V(t,t^{\prime },\chi )e^{i\mathbf{k}\cdot
\Delta \mathbf{x}_{\parallel }}}{c_{1}(u)c_{2}(u)e^{2u}-1}  \notag \\
&&\times \left[ \cosh \left( u\Delta z/z_{0}\right) +\frac{1}{2}\sum_{s=\pm
1}c_{j}^{s}\left( u\right) e^{su|z+z^{\prime }-2z_{j}|/z_{0}}\right] .
\label{Had2b}
\end{eqnarray}%
In the integrand we have defined
\begin{equation}
\chi =b^{-1}\sqrt{u^{2}/z_{0}^{2}-k^{2}}.  \label{xi}
\end{equation}%
The first term in the right-hand side comes from the first integral in (\ref%
{sumfor}). It is further decomposed as%
\begin{eqnarray}
G_{j}(x,x^{\prime }) &=&G_{0}(x,x^{\prime })+\frac{\left( tt^{\prime
}\right) ^{(1-D)/2}}{\left( 2\pi \right) ^{D}b^{D-1}}\int d\mathbf{k}\,e^{i%
\mathbf{k}\cdot \Delta \mathbf{x}_{\parallel }}\int_{0}^{\infty }dy\,  \notag
\\
&&\times \sum_{s=\pm 1}e^{siy\left( z+z^{\prime }-2z_{j}\right) }\frac{%
iy\beta _{j}+s(-1)^{j}}{iy\beta _{j}-s(-1)^{j}}W(t,t^{\prime },\sqrt{%
y^{2}+k^{2}}),  \label{Gj}
\end{eqnarray}%
where%
\begin{equation}
G_{0}(x,x^{\prime })=\frac{\left( b^{2}tt^{\prime }\right) ^{(1-D)/2}}{%
\left( 2\pi \right) ^{D/2}|\Delta \mathbf{x}|^{D/2-1}}\int_{0}^{\infty
}du\,u^{D/2}J_{D/2-1}(u|\Delta \mathbf{x}|)W(t,t^{\prime },u),  \label{G0}
\end{equation}%
with $\Delta \mathbf{x}=(\Delta \mathbf{x}_{\parallel },x^{D}-x^{\prime D})$%
, is the Hadamard function in the geometry (\ref{ds2c}) without boundaries
(for various types of two-point functions in a linearly expanding $D=3$
universe see \cite{Nari80}-\cite{Calz83}).

In the limit $z_{0}\rightarrow \infty $, the second term in the right-hand
side of (\ref{Had2b}) vanishes whereas the term $G_{j}(x,x^{\prime })$
depends on the location $z_{j}$ of a single plate only. From here it follows
that the function $G_{j}(x,x^{\prime })$ corresponds to the Hadamard
function for the geometry of a single plate at $z=z_{j}$. The last term in (%
\ref{Gj}) is the contribution induced by the presence of the plate. It can
be presented in an alternative form rotating the integration contour by the
angle $\pi /2$ for the term with $s=+1$ and by the angle $-\pi /2$ for the
term with $s=-1$. The poles $y=\pm i\sqrt{k^{2}+l^{2}b^{2}}$ on the
imaginary axis are excluded by small semicircles in the right-half plane. In
a way similar to that we have used above, it can be seen that the
contributions from the poles with the upper and lower signs cancel each
other and one gets the representation
\begin{eqnarray}
G_{j}(x,x^{\prime }) &=&G_{0}(x,x^{\prime })+\frac{\left( b^{2}tt^{\prime
}\right) ^{(1-D)/2}}{2\left( 2\pi \right) ^{D-1}b}\int d\mathbf{k}\,e^{i%
\mathbf{k}\cdot \Delta \mathbf{x}_{\parallel }}\int_{k}^{\infty }dy\,  \notag
\\
&&\times \frac{\beta _{j}y+1}{\beta _{j}y-1}e^{-y\left\vert z+z^{\prime
}-2z_{j}\right\vert }V(t,t^{\prime },b^{-1}\sqrt{y^{2}-k^{2}}),  \label{Gj1}
\end{eqnarray}%
where the integral over $y$ is understood in the sense of the principal
value. Substituting this representation into (\ref{Had2b}), the Hadamard
function in the region between two plates is presented in the form%
\begin{eqnarray}
G(x,x^{\prime }) &=&G_{0}(x,x^{\prime })+\frac{\left( b^{2}tt^{\prime
}\right) ^{(1-D)/2}}{\left( 2\pi \right) ^{D-1}bz_{0}}\int d\mathbf{k}%
\,\int_{kz_{0}}^{\infty }du\frac{V(t,t^{\prime },\chi )e^{i\mathbf{k}\cdot
\Delta \mathbf{x}_{\parallel }}}{c_{1}(u)c_{2}(u)e^{2u}-1}  \notag \\
&&\times \left[ \cosh \left( u\Delta z/z_{0}\right) +\frac{1}{2}%
\sum_{j=1,2}c_{j}(u)e^{u|z+z^{\prime }-2z_{j}|/z_{0}}\right] .  \label{G3}
\end{eqnarray}%
In the regions $z<z_{1}$ and $z>z_{2}$, the Hadamard function is given by (%
\ref{Gj1}) with $j=1$ and $j=2$, respectively. The expressions for the
Hadamard functions in the special case $D=1$ are obtained from the formulae
given above omitting the integrations over $\mathbf{k}$ and putting $D=1$, $%
\mathbf{k}=0$.

The explicit extraction of the Hadamard function for the boundary-free
geometry essentially simplifies the renormalization procedure for local
observables at points outside the boundaries. In the vicinity of these
points the local geometry and, hence, the divergences are the same as those
in the corresponding boundary-free problem. As a consequence, the
renormalization is required for the boundary-free contributions only. The
latter procedure for FRW cosmological models is well investigated in the
literature (see, for example, \cite{Birr82B}).

\section{VEV of the field squared}

\label{sec:phi2}

In this and following sections we will investigate the local characteristics
of the vacuum state. As such, first we consider the VEV of the field
squared, denoted here as $\langle 0|\varphi ^{2}|0\rangle \equiv \langle
\varphi ^{2}\rangle $ (in what follows the index $C$ in the notation of the
conformal vacuum state will be omitted). In the region between the plates,
taking the coincidence limit $x^{\prime }\rightarrow x$ in the arguments of
the Hadamard function (\ref{G3}), one gets%
\begin{equation}
\langle \varphi ^{2}\rangle =\langle \varphi ^{2}\rangle _{0}+\langle
\varphi ^{2}\rangle _{\mathrm{b}},  \label{phi2dec}
\end{equation}%
where $\langle \varphi ^{2}\rangle _{0}$ is the renormalized VEV in the
absence of the boundaries and the boundary-induced contribution is given by
the expression%
\begin{equation}
\langle \varphi ^{2}\rangle _{\mathrm{b}}=\frac{B_{D}t^{1-D}}{\left(
z_{0}b\right) ^{D}}\int_{0}^{\infty }dx\,x^{D-2}\,\int_{x}^{\infty }du\frac{%
U(mt,\sqrt{u^{2}-x^{2}}/\left( bz_{0}\right) )}{c_{1}(u)c_{2}(u)e^{2u}-1}%
c(u,z),  \label{phi2n}
\end{equation}%
with the coefficient%
\begin{equation}
B_{D}=\frac{\left( 4\pi \right) ^{(1-D)/2}}{\Gamma ((D-1)/2)}.  \label{BD}
\end{equation}%
In (\ref{phi2n}) and in what follows we use the notations%
\begin{eqnarray}
U(x,y) &=&\frac{J_{y}(x)J_{-y}(x)}{\sin \left( \pi y\right) }  \notag \\
c(u,z) &=&2+\sum_{j=1,2}c_{j}(u)e^{2u|z-z_{j}|/z_{0}}.  \label{cuz}
\end{eqnarray}%
Note that the background geometry is homogeneous and the boundary-free part $%
\langle \varphi ^{2}\rangle _{0}$ does not depend on the spatial point. In
the special case $D=1$, the boundary-induced VEV $\langle \varphi
^{2}\rangle _{\mathrm{b}}$ is obtained from (\ref{phi2n}) omitting $%
B_{D}\int_{0}^{\infty }dx\,x^{D-2}$ and putting in the remaining expression $%
x=0$ and $D=1$.

The boundary-induced contribution in (\ref{phi2n}) is further transformed
passing to a new integration variable $y=\sqrt{u^{2}-x^{2}}$ and introducing
polar coordinates in the plane $(x,y)$. This leads to the result
\begin{equation}
\langle \varphi ^{2}\rangle _{\mathrm{b}}=\frac{B_{D}t^{1-D}}{\left(
bz_{0}\right) ^{D}}\int_{0}^{\infty }du\,\frac{u^{D-1}%
\,S_{D}(mt,u/(bz_{0}))c(u,z)}{c_{1}(u)c_{2}(u)e^{2u}-1},  \label{phi23}
\end{equation}%
with the notation%
\begin{equation}
S_{D}(mt,x)=\int_{0}^{1}ds\,s(1-s^{2})^{(D-3)/2}U(mt,xs).  \label{St}
\end{equation}%
For a massless field
\begin{equation}
S_{D}(mt,x)=\frac{\Gamma ((D-1)/2)}{2\sqrt{\pi }\Gamma (D/2)x},  \label{Sm0}
\end{equation}%
and we can see that the boundary-induced term in (\ref{phi23}) is connected
to the corresponding result in the Minkowski bulk, $\langle \varphi
^{2}\rangle _{\mathrm{b}}^{\mathrm{(M)}}$, by the conformal relation $%
\langle \varphi ^{2}\rangle _{\mathrm{b}}=\left( bt\right) ^{1-D}\langle
\varphi ^{2}\rangle _{\mathrm{b}}^{\mathrm{(M)}}$, where%
\begin{equation}
\langle \varphi ^{2}\rangle _{\mathrm{b}}^{\mathrm{(M)}}=\frac{\left( 4\pi
\right) ^{-D/2}}{\Gamma (D/2)z_{0}^{D-1}}\int_{0}^{\infty }du\,\frac{%
u^{D-2}\,c(u,z)}{c_{1}(u)c_{2}(u)e^{2u}-1}.  \label{phi2M}
\end{equation}

In the regions $z<z_{1}$ and $z>z_{2}$, the VEV of the field squared is
obtained from (\ref{Gj1}). For these regions we have the decomposition%
\begin{equation}
\langle \varphi ^{2}\rangle _{j}=\langle \varphi ^{2}\rangle _{0}+\langle
\varphi ^{2}\rangle _{\mathrm{b}j},  \label{phi2jdec}
\end{equation}%
with the boundary-induced part%
\begin{equation}
\langle \varphi ^{2}\rangle _{\mathrm{b}j}=\frac{B_{D}}{b^{D}t^{D-1}}%
\int_{0}^{\infty }dk\,k^{D-2}\int_{k}^{\infty }dy\,\frac{\beta _{j}y+1}{%
\beta _{j}y-1}e^{-2y\left\vert z-z_{j}\right\vert }U(mt,\sqrt{y^{2}-k^{2}}%
/b).  \label{phi22}
\end{equation}%
Here $j=1$ for the region $z<z_{1}$ and $j=2$ for the region $z>z_{2}$. With
a transformation similar to that used for (\ref{phi23}), the expression (\ref%
{phi22}) can also be presented as%
\begin{equation}
\langle \varphi ^{2}\rangle _{\mathrm{b}j}=\frac{B_{D}}{b^{D}t^{D-1}}%
\int_{0}^{\infty }dy\,y^{D-1}S_{D}(mt,y/b)\frac{\beta _{j}y+1}{\beta _{j}y-1}%
e^{-2y\left\vert z-z_{j}\right\vert }.  \label{phi24}
\end{equation}%
For a massless field, by using (\ref{Sm0}), we obtain the standard relation
with the corresponding result in Minkowski spacetime, $\langle \varphi
^{2}\rangle _{\mathrm{b}j}=\left( bt\right) ^{1-D}\langle \varphi
^{2}\rangle _{\mathrm{b}j}^{\mathrm{(M)}}$, where%
\begin{equation}
\langle \varphi ^{2}\rangle _{\mathrm{b}j}^{\mathrm{(M)}}=\frac{\left( 4\pi
\right) ^{-D/2}}{\Gamma (D/2)}\int_{0}^{\infty }dy\,y^{D-2}\frac{\beta
_{j}y+1}{\beta _{j}y-1}e^{-2y\left\vert z-z_{j}\right\vert },  \label{phi2jM}
\end{equation}%
is the Minkowskian VEV for a massless field.

The boundary-induced contribution (\ref{phi24}) diverges on the boundary $%
z=z_{j}$. For points near the boundary the dominant contribution comes from
large values of $y$. By taking into account that for $u\gg 1$ one has $%
J_{u}(z)J_{-u}(z)/\sin \left( \pi u\right) \sim 1/(\pi u)$, for large $x$ we
get the asymptotic expression%
\begin{equation}
S_{D}(mt,x)\approx \frac{\Gamma ((D-1)/2)}{2\sqrt{\pi }\Gamma (D/2)x}.
\label{SDas}
\end{equation}%
Note that the leading term in the right-hand side coincides with the exact
expression (\ref{Sm0}) for a massless field. Hence, the leading term in the
asymptotic expansion of $\langle \varphi ^{2}\rangle _{\mathrm{b}j}$ for
points near the plates coincides with the corresponding Minkowskian result
multiplied by the conformal factor:
\begin{equation}
\langle \varphi ^{2}\rangle _{\mathrm{b}j}\approx \frac{\left( 4\pi \right)
^{-D/2}\left( 1-2\delta _{0\beta _{j}}\right) }{\Gamma (D/2)\left(
2bt\left\vert z-z_{j}\right\vert \right) ^{D-1}}.  \label{phi2NearPl}
\end{equation}%
Here, for $\beta _{j}\neq 0$, it has also been assumed that $\left\vert
z-z_{j}\right\vert \ll |\beta _{j}|$. The expression on the right of (\ref%
{phi2NearPl}) also gives the leading term near the boundary $z=z_{j}$ for
the VEV\ of the field squared in the region between the boundaries. As seen
from (\ref{phi2NearPl}), near the plates the result for the Neumann boundary
condition is the attractor for the general Robin boundary conditions with $%
\beta _{j}\neq 0$.

Now let us consider the asymptotic of the boundary-induced VEV (\ref{phi24})
at distances from the plate larger than the curvature radius of the
background geometry. This corresponds to the limit $b\left\vert
z-z_{j}\right\vert \gg 1$. The dominant contribution to the integral in (\ref%
{phi24}) comes from the region $y\lesssim 1/\left\vert z-z_{j}\right\vert $
and in this region $y/b\ll 1$. By taking into account that for $u\ll 1$ we
have $J_{u}(z)J_{-u}(z)/\sin \left( \pi u\right) \approx J_{0}^{2}(z)/(\pi
u) $, for small values of $x$ one obtains%
\begin{equation}
S_{D}(mt,x)\approx \frac{\Gamma ((D-1)/2)J_{0}^{2}(mt)}{2\sqrt{\pi }\Gamma
(D/2)x}.  \label{SDas2}
\end{equation}%
With this asymptotic, to the leading order, we get
\begin{equation}
\langle \varphi ^{2}\rangle _{\mathrm{b}j}\approx \frac{J_{0}^{2}(mt)}{%
(bt)^{D-1}}\langle \varphi ^{2}\rangle _{\mathrm{b}j}^{\mathrm{(M)}}.
\label{phi2jlarge}
\end{equation}%
where the Minkowskian VEV\ is given by (\ref{phi2jM}). If in addition $%
\left\vert z-z_{j}\right\vert \gg |\beta _{j}|$, for non-Neumann boundary
conditions the asymptotic takes simpler form%
\begin{equation}
\langle \varphi ^{2}\rangle _{\mathrm{b}j}\approx -\frac{\Gamma
((D-1)/2)J_{0}^{2}(mt)}{\left( 4\pi \right) ^{(D+1)/2}\left( bt\left\vert
z-z_{j}\right\vert \right) ^{D-1}}.  \label{phi2jlarge2}
\end{equation}%
For the Neumann boundary condition the leading term is given by (\ref%
{phi2jlarge2}) with the opposite sign. From (\ref{phi2jlarge2}) it follows
that at large distances from the plate the result for the Dirichlet boundary
condition is the attractor for general Robin boundary conditions. Comparing
with the near-plate asymptotic (\ref{phi2NearPl}), we see that for the Robin
boundary condition, $0<|\beta _{j}|<\infty $, the contribution $\langle
\varphi ^{2}\rangle _{\mathrm{b}j}$ is positive near the plate and negative
at large distances. From (\ref{phi2jlarge2}) it follows that the decay of
the boundary-induced VEV, as a function of the distance from the plate, is
power law for both massless and massive fields. This is in contrast to the
case of the Minkowski bulk, where the decay for a massive field is
exponential, like $e^{-2m\left\vert z-z_{j}\right\vert }$. Under the
condition $b\left\vert z-z_{j}\right\vert \gg 1$ (note that this also
requires the condition $bz_{0}\gg 1$), a relation similar to (\ref%
{phi2jlarge}) is obtained in the region between the plates:%
\begin{equation}
\langle \varphi ^{2}\rangle _{\mathrm{b}}\approx \frac{J_{0}^{2}(mt)}{%
(bt)^{D-1}}\langle \varphi ^{2}\rangle ^{\mathrm{(M)}},  \label{phi2larg}
\end{equation}%
where for $\langle \varphi ^{2}\rangle ^{\mathrm{(M)}}$ we have the
expression (\ref{phi2M}). Note that the dependence on the mass enters in the
argument of the Bessel function only.

\section{VEV of the energy-momentum tensor}

\label{sec:EMT}

In this section we consider the VEV\ of the energy-momentum tensor. It is
expressed in terms of the Hadamard function and the VEV of the field squared
as%
\begin{equation}
\langle T_{\mu \nu }\rangle =\frac{1}{2}\lim_{x^{\prime }\rightarrow
x}\partial _{\mu }\partial _{\nu }^{\prime }G(x,x^{\prime })-\frac{1}{4D}%
\left[ g_{\mu \nu }\nabla _{l}\nabla ^{l}+\left( D-1\right) \left( \nabla
_{\mu }\nabla _{\nu }+R_{\mu \nu }\right) \right] \langle \varphi
^{2}\rangle ,  \label{emtvev1}
\end{equation}%
where the components of the Ricci tensor are given by (\ref{RRik}). Note
that in (\ref{emtvev1}) we have used the expression for the energy-momentum
tensor that differs from the standard one (given, for example, in \cite%
{Birr82B}) by the term vanishing on the solutions of the field equation (see
\cite{Saha04}). The latter will not contribute to the boundary-induced VEV
away from the boundaries. We first consider the VEV in the regions $z<z_{1}$
and $z>z_{2}$. By using the expression (\ref{Gj1}) for the corresponding
Hadamard function, the vacuum energy-momentum tensor is decomposed as%
\begin{equation}
\langle T_{\mu }^{\nu }\rangle _{j}=\langle T_{\mu }^{\nu }\rangle
_{0}+\langle T_{\mu }^{\nu }\rangle _{\mathrm{b}j},  \label{Tnu1pldec}
\end{equation}%
where $\langle T_{\mu }^{\nu }\rangle _{0}$ is the VEV\ in the absence of
boundaries and $\langle T_{\mu }^{\nu }\rangle _{\mathrm{b}j}$ is induced by
the plate at $z=z_{j}$, $j=1,2$.

For the diagonal components of the boundary-induced contribution in (\ref%
{Tnu1pldec}) one gets (no summation over $\mu $)%
\begin{eqnarray}
\langle T_{\mu }^{\mu }\rangle _{\mathrm{b}j} &=&\frac{B_{D}}{b^{D}t^{D+1}}%
\int_{0}^{\infty }dk\,k^{D-2}\int_{k}^{\infty }dy\,\frac{\beta _{j}y+1}{%
\beta _{j}y-1}e^{-2y|z-z_{j}|}  \notag \\
&&\times \left( \hat{f}_{\mu }+\frac{h_{\mu }y^{2}+c_{\mu }k^{2}}{b^{2}}%
\right) U(mt,\sqrt{y^{2}-k^{2}}/b),  \label{Tnubj}
\end{eqnarray}%
where%
\begin{eqnarray}
h_{0} &=&-\frac{D-1}{D},\;h_{l}=\frac{1}{D},\;h_{D}=0,  \notag \\
c_{0} &=&1,\;c_{l}=\frac{1}{1-D},\;c_{D}=0,  \label{cnu}
\end{eqnarray}%
with $l=1,\ldots ,D-1$. The operators in (\ref{Tnubj}) are defined by the
expressions%
\begin{eqnarray}
\hat{f}_{0} &=&\frac{1}{4}\left( t^{2}\partial _{t}^{2}+t\partial
_{t}\right) +t^{2}m^{2},  \notag \\
\hat{f}_{\mu } &=&-\frac{1}{4D}\left( t^{2}\partial _{t}^{2}+t\partial
_{t}\right) ,\;\mu \neq 0.  \label{fnu}
\end{eqnarray}%
Due to the homogeneity of the background spacetime, the boundary-free
contribution $\langle T_{\mu }^{\mu }\rangle _{0}$ does not depend on the
spatial point and the spatial components are isotropic.

The problem under consideration is inhomogeneous along the $t$- and $z$%
-directions. As a consequence of that, in addition to the diagonal
components, the vacuum energy-momentum tensor has a nonzero off-diagonal
component%
\begin{equation}
\langle T_{0}^{D}\rangle _{\mathrm{b}j}=\frac{\mathrm{sgn}(z-z_{j})B_{D}}{%
2D\left( bt\right) ^{D+1}}\int_{0}^{\infty }dk\,k^{D-2}\int_{k}^{\infty
}dy\,y\frac{\beta _{j}y+1}{\beta _{j}y-1}e^{-2y\left\vert z-z_{j}\right\vert
}t\partial _{t}U(mt,\sqrt{y^{2}-k^{2}}/b).  \label{T0D1pl}
\end{equation}%
The corresponding boundary-free part vanishes, $\langle T_{0}^{D}\rangle
_{0}=0$. The off-diagonal component (\ref{T0D1pl}) corresponds to the energy
flux along the direction perpendicular to the plate. It has different signs
in the regions $z<z_{j}$ and $z>z_{j}$. The energy flux can be either
directed from the plate or to the plate. If $\langle T_{0}^{D}\rangle _{%
\mathrm{b}j}>0$ ($\langle T_{0}^{D}\rangle _{\mathrm{b}j}<0$) in the region $%
z>z_{j}$, the energy flux is directed from (to) the plate in both the
regions $z<z_{j}$ and $z>z_{j}$.

In the case of $(1+1)$-dimensional Milne universe ($D=1$) the
boundary-induced VEVs in the geometry of a single plate at $z=z_{j}$ are
given by the expressions (no summation over $\mu $)%
\begin{eqnarray}
\langle T_{\mu }^{\mu }\rangle _{\mathrm{b}j} &=&\frac{1}{bt^{2}}%
\int_{0}^{\infty }dy\,\frac{\beta _{j}y+1}{\beta _{j}y-1}e^{-2y|z-z_{j}|}%
\hat{f}_{\mu }U(mt,y/b),  \notag \\
\langle T_{0}^{D}\rangle _{\mathrm{b}j} &=&\frac{\mathrm{sgn}(z-z_{j})}{%
2\left( bt\right) ^{2}}\int_{0}^{\infty }dy\,y\frac{\beta _{j}y+1}{\beta
_{j}y-1}e^{-2y\left\vert z-z_{j}\right\vert }t\partial _{t}U(mt,y/b).
\label{TD0jD1}
\end{eqnarray}%
In this special case the background geometry is flat and the adiabatic
vacuum coincides with the Minkowskian vacuum.

Alternative expressions for the VEVs in the regions $z<z_{1}$ and $z>z_{2}$,
are obtained in a way we have used for (\ref{phi23}). This gives (no
summation over $\mu $)
\begin{eqnarray}
\langle T_{\mu }^{\mu }\rangle _{\mathrm{b}j} &=&\frac{B_{D}}{b^{D+2}t^{D+1}}%
\int_{0}^{\infty }dy\,y^{D+1}\frac{\beta _{j}y+1}{\beta _{j}y-1}%
e^{-2y|z-z_{j}|}  \notag \\
&&\times \left\{ c_{\mu }S_{D+2}(mt,y/b)+\left[ (b/y)^{2}\hat{f}_{\mu
}+h_{\mu }\right] S_{D}(mt,y/b)\right\} ,  \notag \\
\langle T_{0}^{D}\rangle _{\mathrm{b}j} &=&\frac{\mathrm{sgn}(z-z_{j})B_{D}}{%
2D\left( bt\right) ^{D+1}}\int_{0}^{\infty }dy\,y^{D}\frac{\beta _{j}y+1}{%
\beta _{j}y-1}e^{-2y|z-z_{j}|}t\partial _{t}S_{D}(mt,y/b),  \label{T0D1plNew}
\end{eqnarray}%
with $j=1$ and $j=2$, respectively. For a massless field, by using (\ref{Sm0}%
) and the relation $(D-1)c_{\mu }/D+h_{\mu }=0$, we see that the
boundary-induced VEVs vanish in the regions $z<z_{1}$ and $z>z_{2}$. Of
course, this result could be directly deduced on the base of the conformal
relation with the problem in the Minkowski bulk.

At large distances from the plate, $b\left\vert z-z_{j}\right\vert \gg 1$,
by using the asymptotic expression (\ref{SDas2}), for the diagonal
components we find (no summation over $\mu $)
\begin{eqnarray}
\langle T_{0}^{0}\rangle _{\mathrm{b}j} &\approx &m^{2}\frac{%
J_{0}^{2}(mt)+J_{1}^{2}(mt)}{2\left( bt\right) ^{D-1}}\langle \varphi
^{2}\rangle _{\mathrm{b}j}^{\mathrm{(M)}},  \notag \\
\langle T_{\mu }^{\mu }\rangle _{\mathrm{b}j} &\approx &m^{2}\frac{%
J_{0}^{2}(mt)-J_{1}^{2}(mt)}{2D\left( bt\right) ^{D-1}}\langle \varphi
^{2}\rangle _{\mathrm{b}j}^{\mathrm{(M)}},\;\mu \neq 0,  \label{Tnularge}
\end{eqnarray}%
where $\langle \varphi ^{2}\rangle _{\mathrm{b}j}^{\mathrm{(M)}}$ is given
by (\ref{phi2jM}). For the off-diagonal component the leading term in the
asymptotic expansion has the form%
\begin{equation}
\langle T_{0}^{D}\rangle _{\mathrm{b}j}\approx -\frac{\mathrm{sgn}%
(z-z_{j})mJ_{0}(mt)J_{1}(mt)}{D\left( 4\pi \right) ^{D/2}\Gamma (D/2)\left(
bt\right) ^{D}}\int_{0}^{\infty }dy\,y^{D-1}\frac{\beta _{j}y+1}{\beta
_{j}y-1}e^{-2y|z-z_{j}|}.  \label{T0Dlarge}
\end{equation}%
If additionally $\left\vert z-z_{j}\right\vert \gg |\beta _{j}|$, then in (%
\ref{Tnularge}) one has%
\begin{equation}
\langle \varphi ^{2}\rangle _{\mathrm{b}j}^{\mathrm{(M)}}\approx -\frac{%
\Gamma ((D-1)/2)}{\left( 4\pi \right) ^{(D+1)/2}\left\vert
z-z_{j}\right\vert ^{D-1}}.  \label{phi2Mlarge}
\end{equation}%
For the Dirichlet boundary condition this relation is exact. For the Neumann
boundary condition, $\langle \varphi ^{2}\rangle _{\mathrm{b}j}^{\mathrm{(M)}%
}$ is given by the right-hand side of (\ref{phi2Mlarge}) with the opposite
sign. Under the same conditions, the energy flux decays as $\left\vert
z-z_{j}\right\vert ^{-D}$. In this case the boundary-induced energy density
at large distances is negative for non-Neumann boundary conditions and
positive for the Neumann boundary condition. As regards the vacuum stresses
and the energy flux, they can be either positive or negative depending on
the specific value of $mt$. As we have already emphasized above, at large
distances from the plate the influence of the gravitational field on the
boundary-induced VEVs is essential: for a massive field we have a power law
decay as a function of the distance from the plate, instead of the
exponential suppression in the problem on the Minkowski bulk.

Now let us consider the region between the plates, $z_{1}\leqslant
z\leqslant z_{2}$. By taking into account the expression (\ref{G3}) for the
Hadamard function and using (\ref{emtvev1}), the vacuum energy-momentum
tensor is presented as%
\begin{equation}
\langle T_{\mu }^{\nu }\rangle =\langle T_{\mu }^{\nu }\rangle _{0}+\langle
T_{\mu }^{\nu }\rangle _{\mathrm{b}}.  \label{Tmudec}
\end{equation}%
The diagonal components of the boundary-induced contribution are given by
the formula (no summation over $\mu $)%
\begin{eqnarray}
\langle T_{\mu }^{\mu }\rangle _{\mathrm{b}} &=&\frac{B_{D}}{\left(
bz_{0}\right) ^{D}t^{D+1}}\int_{0}^{\infty }dx\,x^{D-2}\int_{x}^{\infty
}du\left\{ c(u,z)\left[ \hat{f}_{\mu }+\frac{h_{\mu }u^{2}+c_{\mu }x^{2}}{%
(bz_{0})^{2}}\right] \right.   \notag \\
&&\left. -\frac{2d_{\mu }u^{2}}{(bz_{0})^{2}}\right\} \frac{U(mt,\sqrt{%
u^{2}-x^{2}}/\left( bz_{0}\right) )}{c_{1}(u)c_{2}(u)e^{2u}-1},
\label{Tnunu}
\end{eqnarray}%
with $d_{\mu }=1/D$ for $\mu \neq D$ and $d_{D}=-1$. In addition, where is a
nonzero off-diagonal component corresponding to energy flux perpendicular to
the plates:
\begin{eqnarray}
\langle T_{0}^{D}\rangle _{\mathrm{b}} &=&-\frac{B_{D}}{2D(btz_{0})^{D+1}}%
\int_{0}^{\infty }dx\,x^{D-2}\int_{x}^{\infty }du\,u  \notag \\
&&\times \frac{\sum_{j=1,2}\mathrm{sgn}(z-z_{j})c_{j}(u)e^{2u|z-z_{j}|/z_{0}}%
}{c_{1}(u)c_{2}(u)e^{2u}-1}t\partial _{t}U(mt,\sqrt{u^{2}-x^{2}}/\left(
bz_{0}\right) ).  \label{T0D1}
\end{eqnarray}%
If the Robin coefficients for the boundaries are the same, one has $%
c_{1}(u)=c_{2}(u)$. In this special case, the energy flux $\langle
T_{0}^{D}\rangle $ vanishes at $z=(z_{1}+z_{2})/2$ and has opposite signs in
the regions $z<(z_{1}+z_{2})/2$ and $z>(z_{1}+z_{2})/2$.

In the case $D=1$, the VEV of the energy-momentum tensor is given by the
expressions (no summation over $\mu $)%
\begin{eqnarray}
\langle T_{\mu }^{\mu }\rangle _{\mathrm{b}} &=&\frac{1}{bz_{0}t^{2}}%
\int_{0}^{\infty }du\,\left[ c(u,z)\hat{f}_{\mu }-\frac{2(-1)^{\mu }}{%
(bz_{0})^{2}}u^{2}\right] \frac{U(mt,u/\left( bz_{0}\right) )}{%
c_{1}(u)c_{2}(u)e^{2u}-1},  \notag \\
\langle T_{0}^{1}\rangle _{\mathrm{b}} &=&-\frac{1}{2(bz_{0}t)^{2}}%
\int_{0}^{\infty }du\,u\frac{\sum_{j=1,2}\mathrm{sgn}%
(z-z_{j})c_{j}(u)e^{2u|z-z_{j}|/z_{0}}}{c_{1}(u)c_{2}(u)e^{2u}-1}t\partial
_{t}U(mt,u/\left( bz_{0}\right) ),  \label{T01D1}
\end{eqnarray}%
with $\mu =0,1$.

Introducing in (\ref{Tnunu}) and (\ref{T0D1}) a new integration variable $y=%
\sqrt{u^{2}-x^{2}}$ and passing to polar coordinates in the $(x,y)$-plane,
we obtain equivalent representations (no summation over $\mu $)
\begin{eqnarray}
\langle T_{\mu }^{\mu }\rangle _{\mathrm{b}} &=&\frac{B_{D}}{\left(
z_{0}b\right) ^{D+2}t^{D+1}}\int_{0}^{\infty }du\,\frac{u^{D+1}}{%
c_{1}(u)c_{2}(u)e^{2u}-1}\left\{ c_{\mu }c(u,z)S_{D+2}(mt,u/(bz_{0}))\right.
\notag \\
&&\left. +\left[ c(u,z)\left( (bz_{0}/u)^{2}\hat{f}_{\mu }+h_{\mu }\right)
-2d_{\mu }\right] S_{D}(mt,u/(bz_{0}))\right\} ,  \label{TnuNew}
\end{eqnarray}%
for the diagonal components and
\begin{equation}
\langle T_{0}^{D}\rangle =-\frac{B_{D}}{2D(btz_{0})^{D+1}}\int_{0}^{\infty
}du\,u^{D}\frac{\sum_{j=1,2}\mathrm{sgn}%
(z-z_{j})c_{j}(u)e^{2u|z-z_{j}|/z_{0}}}{c_{1}(u)c_{2}(u)e^{2u}-1}t\partial
_{t}S_{D}(mt,u/(bz_{0})),  \label{T0DNew}
\end{equation}%
for the off-diagonal component. In the case of a massless field, by taking
into account (\ref{Sm0}), one gets (no summation over $\mu $)%
\begin{equation}
\langle T_{\mu }^{\mu }\rangle _{\mathrm{b}}=-\frac{2\left( 4\pi \right)
^{-D/2}d_{\mu }}{\Gamma (D/2)\left( z_{0}bt\right) ^{D+1}}\int_{0}^{\infty
}du\,\frac{u^{D}}{c_{1}(u)c_{2}(u)e^{2u}-1},  \label{Tnum0}
\end{equation}%
and the off-diagonal component vanishes. In this case we have a conformal
relation with the corresponding problem in the Minkowski bulk. For a massive
field the VEVs (\ref{TnuNew}) and (\ref{T0DNew}) diverge on the plates. The
divergences on the plate at $z=z_{j}$ are the same as those for $\langle
T_{\mu }^{\nu }\rangle _{\mathrm{b}j}$. The part in the VEV induced by the
presence of the second plate, $\langle T_{\mu }^{\nu }\rangle _{\mathrm{b}%
}-\langle T_{\mu }^{\nu }\rangle _{\mathrm{b}j}$, is finite on the first
plate.

By using the relations $\sum_{\mu =0}^{D}\hat{f}_{\mu }=t^{2}m^{2}$ and $%
\sum_{\mu =0}^{D}h_{\mu }=\sum_{\mu =0}^{D}d_{\mu }=0$, we can check that
the boundary-induced contributions in all the regions obey the trace
relation
\begin{equation}
\langle T_{\mu }^{\mu }\rangle _{\mathrm{b}}=m^{2}\langle \varphi
^{2}\rangle _{\mathrm{b}}.  \label{TrRel}
\end{equation}%
For a massless field the boundary-induced contribution in the VEV of the
energy-momentum tensor is traceless. The trace anomaly is contained in the
boundary-free part only. As an additional check, we can see that the
boundary-induced VEVs satisfy the covariant conservation equation $\nabla
_{\mu }\langle T_{\nu }^{\mu }\rangle _{\mathrm{b}}=0$. For the geometry
under consideration it is reduced to the following two equations%
\begin{eqnarray}
t^{-D}\partial _{t}\left( t^{D+1}\langle T_{0}^{0}\rangle _{\mathrm{b}%
}\right) +\frac{1}{b}\partial _{z}\langle T_{0}^{D}\rangle _{\mathrm{b}%
}-\langle T_{\mu }^{\mu }\rangle _{\mathrm{b}} &=&0,  \notag \\
t^{-D}\partial _{t}\left( t^{D+1}\langle T_{0}^{D}\rangle _{\mathrm{b}%
}\right) -\frac{1}{b}\partial _{z}\langle T_{D}^{D}\rangle _{\mathrm{b}}
&=&0.  \label{ContEq}
\end{eqnarray}%
The second of these equations shows that the inhomogeneity of the normal
stress is related to the nonzero energy flux along the direction normal to
the plates.

Note that we have considered the components of the vacuum energy-momentum
tensor in the coordinate system $(\eta ,x^{1},x^{2},\ldots ,x^{D})$. In the
coordinate system with the proper time $t$, $(t,x^{1},x^{2},\ldots ,x^{D})$,
the diagonal components $\langle \tilde{T}_{\mu }^{\mu }\rangle $ are the
same, $\langle \tilde{T}_{\mu }^{\mu }\rangle =\langle T_{\mu }^{\mu
}\rangle $, whereas for the off-diagonal component one has $\langle \tilde{T}%
_{0}^{D}\rangle =\langle T_{0}^{D}\rangle /(bt)$.

\section{The Casimir forces}

\label{sec:Force}

The vacuum force acting per unit surface of the plate at $z=z_{j}$ is
determined by the normal stress $\langle T_{D}^{D}\rangle |_{z=z_{j}}$. For
a massive field this quantity diverges. The divergence comes from the single
plate contribution $\langle T_{D}^{D}\rangle _{\mathrm{b}j}$. The latter is
the same on the left- and right-hand sides of the plate and, hence, the
corresponding net force is zero. The same is the case for the boundary-free
part $\langle T_{D}^{D}\rangle _{0}$. Consequently, the resulting force
comes from the second plate-induced part $\langle T_{D}^{D}\rangle -\langle
T_{D}^{D}\rangle _{j}$ and the corresponding effective pressure is given by $%
P_{j}=\left( \langle T_{D}^{D}\rangle _{j}-\langle T_{D}^{D}\rangle \right)
|_{z=z_{j}}$, where $\langle T_{D}^{D}\rangle $ is the normal stress in the
region between the plates. The forces corresponding to $P_{j}$ act on
the sides of $z=z_{1}+0$ and $z=z_{2}-0$ of the plates. They are attractive
(repulsive) for negative (positive) $P_{j}$. By taking into account the
expressions (\ref{Tnubj}) and (\ref{Tnunu}), the vacuum pressures on the
plates are presented as%
\begin{eqnarray}
P_{j} &=&-B_{D}\frac{\left( bz_{0}\right) ^{-D}}{t^{D+1}}\int_{0}^{\infty
}dx\,x^{D-2}\int_{x}^{\infty }du\,\frac{2\left( u/bz_{0}\right) ^{2}+\left[
2+c_{j}(u)+1/c_{j}(u)\right] \hat{f}_{D}}{c_{1}(u)c_{2}(u)e^{2u}-1}  \notag
\\
&&\times U(mt,\sqrt{u^{2}-x^{2}}/(bz_{0})),  \label{Pjnew}
\end{eqnarray}%
with $\hat{f}_{D}$ given by (\ref{fnu}). As before, the integrals are
understood in the sense of the principal value. Depending on the Robin
coefficients and on the value of $mt$, the forces corresponding to (\ref%
{Pjnew}) can be either attractive or repulsive.

An alternative expressions for the forces acting on the plates are obtained
by using the normal stresses from (\ref{T0D1plNew}) and
(\ref{TnuNew}):%
\begin{equation}
P_{j}=-\frac{B_{D}}{\left( z_{0}b\right) ^{D}t^{D+1}}\int_{0}^{\infty
}du\,u^{D-1}\frac{2\left( u/bz_{0}\right) ^{2}+\left[ 2+c_{j}(u)+1/c_{j}(u)%
\right] \hat{f}_{D}}{c_{1}(u)c_{2}(u)e^{2u}-1}S_{D}(mt,u/(bz_{0})).
\label{Pjnew2}
\end{equation}%
In particular, one can have the situation when the forces are repulsive at
small separations between the plates and attractive at large separations.
For a massless field, by using the expression (\ref{Sm0}) for the function\ $%
S_{D}(mt,x)$, one gets $P_{j}=P_{j}^{\mathrm{(M)}}/\left( bt\right) ^{D+1}$,
where
\begin{equation}
P_{j}^{\mathrm{(M)}}=-\frac{2\left( 4\pi \right) ^{-D/2}}{\Gamma
(D/2)z_{0}^{D+1}}\int_{0}^{\infty }du\,\frac{u^{D}}{c_{1}(u)c_{2}(u)e^{2u}-1}%
,  \label{Pjnew2m0}
\end{equation}%
is the corresponding pressure for plates in the Minkowski bulk with the
separation $z_{0}$. Note that in the problem under consideration $z_{0}bt$
is the proper distance between the plates for a fixed $t$. For the Minkowski
bulk the Casimir forces are the same for separate plates, independently on
the values of the Robin coefficient. As seen from (\ref{Pjnew2m0}), in
general, this is not the case for an expanding universe.

At small separations between the plates, compared with the curvature radius
of the background spacetime, one has $bz_{0}\ll 1$. By using the asymptotic
expression (\ref{SDas}) for the function $S_{D}(mt,x)$, to the leading order
one gets $P_{j}\approx P_{j}^{\mathrm{(M)}}/\left( bt\right) ^{D+1}$, where $%
P_{j}^{\mathrm{(M)}}$ is given by (\ref{Pjnew2m0}). In the limit under
consideration the effects of gravity on the Casimir forces are small and the
leading term coincides with that in the Minkowski bulk multiplied by the
conformal factor. If in addition $z_{0}\ll |\beta _{j}|$, the leading term
is further simplified as
\begin{equation}
P_{j}\approx -\frac{D\zeta (D+1)\Gamma (\left( D+1\right) /2)}{\left( 4\pi
\right) ^{(D+1)/2}\left( z_{0}bt\right) ^{D+1}},  \label{Pjsmall}
\end{equation}%
with $\zeta (x)$ being the Riemann zeta function. The same leading term 
is obtained for Dirichlet boundary conditions ($\beta _{j}=0$). The corresponding 
forces are attractive. For the Dirichlet
boundary condition on one plate and for non-Dirichlet boundary condition on
the other the forces are repulsive at small separations.

For the separation between the plates larger than the curvature radius, $%
bz_{0}\gg 1$, for the function $S_{D}(mt,x)$ in the integrand of (\ref%
{Pjnew2}) we use asymptotic (\ref{SDas2}). To the leading order, for
non-Dirichlet boundary conditions ($\beta _{j}\neq 0$) one gets%
\begin{equation}
P_{j}\approx \frac{m^{2}\left[ J_{1}^{2}(mt)-J_{0}^{2}(mt)\right] }{2D\left(
4\pi \right) ^{D/2}\Gamma (D/2)\left( btz_{0}\right) ^{D-1}}\int_{0}^{\infty
}du\,u^{D-2}\frac{2+c_{j}(u)+1/c_{j}(u)}{c_{1}(u)c_{2}(u)e^{2u}-1}.
\label{Pjas}
\end{equation}%
In particular, for the Neumann boundary condition we find%
\begin{equation}
P_{j}\approx \frac{2m^{2}\Gamma (\left( D-1\right) /2)\zeta (D-1)}{D\left(
4\pi \right) ^{(D+1)/2}\left( btz_{0}\right) ^{D-1}}\left[
J_{1}^{2}(mt)-J_{0}^{2}(mt)\right] .  \label{PjasN}
\end{equation}%
The corresponding Casimir forces can be either attractive or repulsive. For
a massless field the leading terms (\ref{Pjas}) and (\ref{PjasN}) vanish.
For the Dirichlet boundary condition on both the plates one has $c_{j}(u)=-1$
and the leading term is given by%
\begin{equation}
P_{j}\approx -\frac{D\Gamma (\left( D+1\right) /2)\zeta (D+1)}{\left( 4\pi
\right) ^{(D+1)/2}\left( btz_{0}\right) ^{D+1}}J_{0}^{2}(mt).  \label{PjasD}
\end{equation}%
In this case the forces are attractive. Note that for plates in the
Minkowski bulk the Casimir forces are attractive for both the Dirichlet and
Neumann boundary conditions at all separations between the plates.

\section{Conclusion}

\label{sec:Conc}

We have investigated combined effects of the background gravitational field
and boundaries on the quantum properties of the scalar vacuum. As a
background geometry a linearly expanding spatially flat universe is taken.
In a special case with a single spatial dimension the geometry is flat and
coincides with the Milne universe. The boundary geometry is given by two
parallel plates on which the field obeys the Robin boundary conditions with
the coefficients being linear functions of the proper time coordinate $t$.
We have shown that, with this dependence, the problem is exactly solvable.
The two-point functions, describing all the properties of the quantum vacuum
in the model under consideration, are presented in the form of the mode-sum
over a complete set of scalar modes obeying the boundary conditions. These
modes are given by (\ref{phi+}) with the time dependence defined by (\ref{fM}%
) or, equivalently, by (\ref{fJ}). These functions contain an arbitrary
constant which is fixed by the choice of the vacuum state. We have
considered two special cases corresponding to the adiabatic and conformal
vacua.

The evaluation of the VEVs is presented for the example of a conformally
coupled scalar field in the conformal vacuum state. In the region between
the plates the corresponding Hadamard function is given by the expression (%
\ref{Had2b}). In that representation, $G_{j}(x,x^{\prime })$ is the Hadamard
function in the geometry of a single plate at $z=z_{j}$. It is further
decomposed into the boundary-free and boundary-induced contributions, given
by (\ref{Gj1}). The two-point functions in the regions $z<z_{1}$ and $%
z>z_{2} $ have the form (\ref{Gj1}) for $j=1$ and $j=2$, respectively. With
the explicitly extracted boundary-free part in the Hadamard function, for
points away from the boundaries, the renormalization of the local VEVs in
the coincidence limit is reduced to the renormalization in the boundary-free
geometry. The latter procedure is well investigated in the literature for
general Friedmann-Robertson-Walker cosmological models and we were mainly
concerned with the boundary-induced effects.

As an important local characteristic of the vacuum state, we have firstly
considered the VEV\ of the field squared. Two equivalent representations for
the boundary-induced contribution, (\ref{phi2n}) and (\ref{phi23}), have
been provided in the region between the plates. Similar representation for
the regions $z<z_{1}$ and $z>z_{2}$ have the form (\ref{phi22}) and (\ref%
{phi24}). For a massless field, the boundary-induced VEVs are connected with
the corresponding VEVs in the Minkowski bulk by the standard conformal
relation. For points near the plates, the dominant contribution to the VEVs
comes from the fluctuations with short wavelengths and the effects of
gravity on the boundary-induced VEVs are weak. The influence of the
gravitational field is essential at distance from the plates large than the
curvature radius of the background spacetime. In the geometry of a single
plate the leading term in the corresponding asymptotic expansion is given by
(\ref{phi2jlarge2}), where $\langle \varphi ^{2}\rangle _{\mathrm{b}j}^{%
\mathrm{(M)}}$ is the corresponding VEV for a massless field in the
Minkowski bulk. In contrast to the latter geometry, for massive fields the
decay of the boundary-induced VEV in the problem at hand is power law.

Among the physical quantities playing a central role in quantum field theory
on curved spacetime is the VEV of the energy-momentum tensor. Similar to the
VEV\ of the field squared, it is decomposed into the boundary-free and
boundary-induced parts. As a consequence of the time dependence of the
background geometry, the boundary-induced contribution has a nonzero
off-diagonal component corresponding to the energy flux along the direction
normal to the boundaries. In the regions $z<z_{1}$ and $z>z_{2}$ the latter
is presented in two equivalent forms, given by (\ref{Tnubj}), (\ref{T0D1pl})
and (\ref{T0D1plNew}). The effects of the gravity are crucial at distances
larger than the curvature radius. The corresponding asymptotics are given by
(\ref{Tnularge}) and (\ref{T0Dlarge}) and the decay of the boundary-induced
contributions, as functions of the distance from the plate, is power law for
both massless and massive fields. In the region between the plates the
corresponding components are presented by the formulas (\ref{Tnunu}), (\ref%
{T0D1}), (\ref{TnuNew}) and (\ref{T0DNew}). We have explicitly shown that
the boundary-induced contributions obey the trace relation (\ref{TrRel}) and
the covariant conservation equation. The latter is reduced to the equations (%
\ref{ContEq}). For a massless filed the problem under consideration is
conformally related to the corresponding problem in the Minkowski bulk. In
this special case the off-diagonal component vanishes and the
boundary-induced contribution in the VEV\ of the energy-momentum tensor is
traceless. The trace anomaly is present in the boundary-free part only.

We have also investigated the Casimir forces. The vacuum pressure on the
plates is decomposed into the self action and interaction contributions. The
latter is induced by the presence of the second plate. Because of the
homogeneity of the background spacetime, the self action parts are the same
on the left- and right-hand sides of the plates. As a consequence, the
corresponding net force becomes zero and the Casimir forces are conditioned
by the presence of the second plate. The force per unit surface acting on
the plate at $z=z_{j}$ is given by the expressions (\ref{Pjnew}) and (\ref%
{Pjnew2}). Unlike to the problem in the Minkowski bulk, for a massive field
the Casimir force acting on the left and right plates are different if the
Robin coefficients differ. At large separations between the plates, compared
with the curvature radius, the leading term in the asymptotic expansion of
the Casimir pressure is given by (\ref{Pjas}). In the special case of the
Neumann boundary condition it takes simpler form (\ref{PjasN}) and, unlike
to the Minkowskian geometry, the corresponding forces can be either
repulsive or attractive. For the Dirichlet boundary condition the leading
term vanishes and the next-to-leading term is presented as (\ref{PjasD}).
The corresponding forces are attractive.

\end{document}